\documentclass[preprint, authoryear]{elsarticle}
\usepackage[utf8]{inputenc}
\usepackage{amsmath}
\usepackage{amsfonts}
\usepackage{amssymb}
\usepackage{amsthm}
\usepackage{epsf}
\usepackage{array}
\usepackage{graphicx}
\usepackage{color}
\usepackage{bm}
\usepackage{bbm}
\usepackage{url}
\usepackage{color}
\usepackage{natbib}
\usepackage{setspace}

\newtheorem{lemma}{Lemma}

\newtheorem{theorem}{Theorem}
\newtheorem{proposition}{Proposition}
{
\theoremstyle{definition}
\newtheorem{remark}{Remark}

}

\def\cL{\mathcal{L}^*}
\def\calL{\mathcal{L}}
\def\Va{\bar{V}}
\def\hv{\hat{V}}
\def\ha{\hat{a}}
\def\ba{\bar{a}}
\def\sa{a^*}
\def\Vb{V}

\def\spacingset#1{\renewcommand{\baselinestretch}{#1}\small\normalsize} 
\def\hX{\hat{X}}
\def\hZ{\hat{Z}}
\def\brt{\tau_0^{\hat{X}}}

\def\cF{\mathcal{F}}
\def\sP{\mathsf{P}}
\def\cA{\mathcal{A}}

\def\sE{\mathsf{E}}
\def\BP{M}
\def\aopt{\hat{a}}

\title{Fiscal stimulus as an optimal control problem}
 
\author[rice]{Philip~A.~Ernst}
\ead{philip.ernst@rice.edu}

\author[imerman]{Michael~B.~Imerman} 
\ead{mbi212@lehigh.edu}

\author[shepp]{Larry~Shepp}

\author[tamu]{Quan~Zhou}
\ead{quan.zhou@rice.edu}

\journal{SPA Special Issue: ``In Memoriam: Larry Shepp''}

\address[rice]{Department of Statistics, Rice University}
\address[imerman]{Drucker School of Management, Claremont Graduate University}
\address[shepp]{Deceased April 23, 2013}
\address[tamu]{Department of Statistics, Texas A\&M University}

\begin{document}

\begin{abstract}
During the Great Recession, Democrats in the United States argued that government spending could be utilized to ``grease the wheels'' of the economy in order to create wealth and to increase employment; Republicans, on the other hand, contended that government spending is wasteful and discouraged investment, thereby increasing unemployment.  
Today, in 2020, we find ourselves in the midst of another crisis where government spending and fiscal stimulus is again being considered as a solution.
In the present paper, we address this question by formulating an optimal control problem generalizing the model of  \citet{rs}. The model allows for the company to borrow continuously from the government. 
We prove that there exists an optimal strategy; rigorous verification proofs for its optimality are provided. We proceed to prove that government loans increase the expected net profit of a company. We also examine the consequences of different profit-taking behaviors among firms who receive fiscal stimulus.

\begin{keyword} Dividend problem \sep Radner\textendash Shepp model  \sep financial stimulus \sep Hamilton-Jacobi-Bellman equation \sep stochastic control.

\MSC[2010] Primary: 60H10, 60J60 \sep Secondary: 60G15
\end{keyword}

\end{abstract}

\maketitle
\vfill
\spacingset{1.1}
\newpage

\section{Introduction}

The purpose of this paper  is to mathematically model optimal fiscal policy with the hopes of contributing to the ongoing debate between United State Democrats and Republicans on what government measures should be taken to stimulate the economy.  The model was originally developed by the second named and the third named authors in 2010, shortly after the Great Recession and the U.S. government's subsequent bailout of large banks. With the present risk of a major global recession from the COVID-19 pandemic, this work offers a timely addition to the third named author's seminal contributions to mathematical finance and optimal control.\footnote{This paper has been written for the special issue in memory of the third author, Larry Shepp, who passed away on April 23rd, 2013. During his later years, one of Shepp's key research interests concerned stochastic modelling in finance and economics.}

The key macroeconomic question considered in this paper is how much -- if at all -- should the government inject into the private sector to improve aggregate wealth of the economy?  To this end,  we assume that a company can be characterized by four parameters, $(x,\mu,\sigma,r)$: the present cash reserve, $x > 0$, the profit rate, $\mu > 0$, the riskiness, $\sigma > 0$, and the prevailing interest rate, $r \geq 0$.  Let $(\Omega, \cF, \{\cF_t\}_{t \geq 0}, \sP)$ be a filtered probability space, where $\cF_t$ represents the market information available at time $t$.
Let  $W(t)$ be a Wiener process  w.r.t. $\{\cF_t\}_{t \geq 0}$, representing uncertainty, and $Z(t)$ be the total dividends subtracted from the fortune up until time $t$.
The cash reserve of the company at time $t$, denoted by $X(t)$, evolves according to the dynamics
\begin{equation}\label{eq:def.X}
X(t) = x + \mu t + \sigma W(t) - Z(t), \quad t \ge 0, 
\end{equation}
where $W(0) =  0$ and $Z(0) \geq 0$. 

To model fiscal policies that can ``grease the wheels" of the private sector, we assume that the government may choose to provide a loan to the firm in order to increase the firm's expected profit rate from $\mu$ to $\mu^*$, where it is assumed that $\mu^* \geq \mu > 0$. This loan is to be repaid at the interest rate $r$.
It is assumed that the firm may borrow continuously at a limited rate from the government.\footnote{It has been shown in the macroeconomics literature that government loans and subsidies to private firms can be an effective means of fiscal stimulus~\citep{lucas}.}  The above considerations lead us to write the process $Z(t)$ as $Z(t) = Z_+(t) - Z_-(t)$ where $Z_+(t)$ and $Z_-(t)$, respectively, represent the total dividends (which may be taxed) issued up to time $t$ and the total loans issued up to time $t$. Hence, we require $Z_\pm(t) \geq 0$ for any $t$ and rewrite equation~\eqref{eq:def.X} as 
\begin{equation}\label{eq:def.XX}
X(t) = x + \mu t + \sigma W(t) + Z_-(t) - Z_+(t).
\end{equation}
Let $\cA$ be the collection of admissible controls $(Z_+, \, Z_-)$ which satisfy assumptions (A1)-(A3) below:
\begin{enumerate}[({A}1)]
\item $Z_+(t)$ is a nondecreasing and c\`{a}dl\`{a}g   process adapted to $\{\cF_t\}_{t \geq 0}$.  
\item $Z_+(0) \in [0, x]$, and $\Delta Z_+(t)   =  Z_+(t) - Z_+(t-)\geq X(t -)$  for every $t > 0$.
\item $Z_-(t)$ is a nondecreasing and differentiable process adapted to $\{\cF_t\}_{t \geq 0}$ with $Z_-(0) = 0$ and  $ d Z_-(t) / dt \leq \mu^* - \mu$ for some $\mu^* \geq \mu$. 
\end{enumerate}
Condition (A2) implies that the company cannot make a lump-sum dividend payment greater than its current fortune.  
Condition (A3) means that the company may be supplied any amount less than or equal to $(\mu^* - \mu) dt$ in each interval $dt$.
It is worth noting that the restriction we put on $Z_-(t)$  is different from the usual setup of dividend problems where singular controls are used to represent instantaneous large-scale capital injections.  
We require that the government loan is paid back at interest rate $r$, and the objective of the firm is to maximize its expected net value (to be defined) by choosing the policy $(Z_+,  Z_-)$ optimally. Mathematically, this means we want to compute the value function defined as\footnote{
Throughout this paper, we will frequently use notations such as $\Vb(x; \mu, \mu^*)$ to indicate that the value function depends on the drift parameters. 
But parameters $\sigma$ and $r$ are always omitted since they are not of direct interest and can be treated as fixed. 
} 
\begin{equation}\label{eq:def.Vb}
  \Vb(x) = \Vb(x; \mu, \mu^*) = \sup_{ (Z_+, Z_-) \in \cA} \sE_x    \int_{0-}^{\tau_0} e^{-rt} [ dZ_+(t) - dZ_-(t) ], 
\end{equation} 
where $\tau_0  = \inf\{t:  \,  X(t) \leq 0 \}$ is the time that the company goes bankrupt and the notation $\int_{0-}^\tau d Z(t)$  should be understood as $Z(0) + \int_0^\tau d Z(t)$. The definition implies that $V(0) = 0$.  Further,  $V(x) \ge x$ for any $x$ since one strategy for taking profit is to take the fortune immediately; bankruptcy occurs at time $\tau_0 = 0$.  

If we replace $\cA$ with a smaller admissible class $\cA' = \{  (Z_+, \, Z_-) \in \cA: \, Z_-(t) \equiv 0 \}$ and define the corresponding value function by 
\begin{equation}\label{eq:def.V1}
\Va(x) = \sup_{ (Z_+, Z_-) \in \cA'} \sE_x    \int_{0-}^{\tau_0} e^{-rt}  dZ_+(t), 
\end{equation} 
the problem then reduces to that considered in the seminal paper of~\cite{rs}.   But this is equivalent to considering the function $\Vb(x; \mu, \mu)$, which corresponds to the case where the maximum loan rate is zero.  
Since $\cA' \subset \cA$, we have
\begin{equation}\label{eq:vs.ineq}
 \Vb(x; \mu,\mu^* )  \geq 
  \Vb(x; \mu,  \mu ) = \Va(x). 
\end{equation}
The question is as follows: if $\mu^* > \mu$, whether $Z_-(t) = 0$ is strictly suboptimal, i.e., whether we have $\Vb(x) > \Va(x)$?
It turns out that we do indeed have strict inequality, which implies that the expected additional dividend payouts from having the government funds is strictly greater than the loan cost (until bankruptcy), provided that the company takes profits  in an optimal way to maximize its presumed objective. More generally, we have $\Vb(x; \mu, \mu^* ) > \Vb(x; \mu,\mu' )$ for any $\mu^* > \mu' \geq  \mu$; that is, the more the fiscal stimulus offered by the government, the larger net value of the company (see Section~\ref{sec4}.) 
  
Now, what if the company, after borrowing from the government, chooses a ``greedy" policy that maximizes its own dividend payouts without caring to repay the loans?  In Section~\ref{sec:disc.greedy}, we will show that such a strategy is socially undesirable in that the expected net value of the company could be smaller than that with no government loans and, moreover, the expected dividend payouts may not even cover the loan cost.  
This represents an interesting caveat to the results of our model: in order to ensure that the mathematically optimal and socially optimal solution is achieved, the government must play some role in enforcing how the firms who take government money operate.  
 
The paper is organized as follows. In Section~\ref{sec:rs} we review the related literature and, in particular, the results of the seminal work of~\cite{rs}, which can be viewed as a baseline model where the government does not offer any loans to companies. In Section~\ref{sec4} we derive the corresponding free-boundary problem for the value function given in equation~\eqref{eq:def.Vb} and prove the existence of the solution. 
Further optimal control results for our problem are provided in Section~\ref{sec:barrier}, including how the optimal dividend payout policy changes with the model parameters. 
In Section~\ref{sec:disc}, we discuss the economic implications of different dividend payout policies. 
Section~\ref{sec:proofs} concludes the paper with the requisite technical proofs.

\section{Preliminaries}\label{sec:rs}
  
\subsection{Solution to the Radner--Shepp model}
Consider the problem with $\mu^* = \mu$ and the value function $\Va(x)$ defined in equation~\eqref{eq:def.V1}, which we shall refer to as the Radner--Shepp  model~\citep{rs}. 
The solution  was found by~\cite{rd},~\cite{jeanblanc1995optimization} and~\cite{asmussen1997controlled}, and~\cite{rd} further showed that the company that follows the optimal policy will go bankrupt in a finite time with probability $1$.  
The optimal policy is to pay out dividends at a reflection barrier $\ba$: if  $X(t) \leq  \ba$, no payment is made; otherwise, an instantaneous payment is made so that $X(t)$ drops to $\ba$. 
By solving the Hamilton-Jacobi-Bellman equation
\begin{align*} 
\max\{\calL v (x),  1 - v'(x) \} = 0,     \quad \quad 
\calL = -r  + \mu \dfrac{\partial}{\partial x} + \dfrac{\sigma^2}{2} \dfrac{\partial^2}{\partial x^2}, 
\end{align*}
with the  initial condition $v (0) = 0$, we obtain the solution 
\begin{equation}\label{eq:barV}
\Va(x) = \left\{\begin{array}{cc}
 \dfrac{e^{\gamma_+ x} - e^{\gamma_- x} }{\gamma_+ e^{\gamma_+ \ba} - \gamma_- e^{\gamma_- \ba} },  & \quad 0 \le x \le \ba, \medskip \\
\Va(\ba) + (x- \ba), & \quad \ba < x < \infty, 
\end{array}
\right.
\end{equation}
where $\gamma_+, \gamma_-$ are  the roots of the indicial equation,  
$\gamma_\pm =  ( -\mu \pm \sqrt{\mu^2 + 2r\sigma^2} )/ \sigma^2.$
The unknown optimal threshold $\ba$ can be most  easily determined by the smooth-fit heuristic, $\Va^{\prime \prime}(\ba) = 0$, 
which yields 
\begin{equation}\label{eq:def.a1}
\ba = \frac{2}{\gamma_+ - \gamma_-} \log{ \left| \frac{\gamma_-}{\gamma_+} \right| }, 
\end{equation}
given $\mu \geq 0$. If $\mu < 0$,  then one can show $\ba = 0$ and $\Va(x) = x$;  that is, it is always optimal to ``take the money and run.''  Without loss of generality, henceforth we assume $\mu > 0$, which means the company is {\em profitable}. 
Interestingly, the threshold $\ba$ goes to zero either as $\mu \downarrow 0$ or as $\mu \uparrow \infty$.  
We also point out for further reference that
\begin{equation}\label{eq:va}
\Va(\ba) = \frac{e^{\gamma_+ \ba} - e^{\gamma_- \ba} }{\gamma_+ e^{\gamma_+ \ba} - \gamma_- e^{\gamma_- \ba} } = \frac{\gamma_+ + \gamma_-}{\gamma_+ \gamma_-} = \frac{\mu}{r}. 
\end{equation}

To study the optimal payout policy, we introduce the notation $\BP_a(t; \mu)$, which denotes the the unique solution to the  Skorokhod reflection problem of the process $x + \mu t + \sigma W(t)$ reflected at the barrier $a > 0$~\citep{watanabe1971stochastic, tanaka1979}. That is,  
\begin{equation}\label{eq:def.Za}
\BP_a(t; \mu) =  \sup_{0 \leq s \leq t}  ( x + \mu s + \sigma W(s) - a )^+, 
\end{equation}
where $w^+ = w \vee 0$ denotes the positive part of $w$.  The supremum in the equation~\eqref{eq:def.V1} is attained at $Z_+(t) = \BP_{\ba}(t; \mu)$~\citep{rs}.

For an arbitrary barrier payout policy with level $a$,  we define the expected time-discounted total dividend payouts as\footnote{We write $\tau_0 = \tau_0(a, \mu)$ to indicate that the bankruptcy time depends on $\mu$ and $a$.}
\begin{equation}\label{eq:def.D}
D_a(x) = D_a(x; \mu) =  \sE_x \, \left[ \BP_a(0; \mu)  +\int_0^{\tau_0(a, \mu)} e^{-rt} d\BP_a(t; \mu) \right]. 
\end{equation} 
As shown in \citet{shreve1984optimal},   the formula in equation~\eqref{eq:barV} is still applicable with $\ba$ replaced by $a$, from which we obtain 
\begin{equation}\label{eq:Da}
D_a(x; \mu ) =  \left\{\begin{array}{cc}
  \dfrac{e^{\gamma_+ x} - e^{\gamma_- x} }{\gamma_+ e^{\gamma_+ a} - \gamma_- e^{\gamma_- a} } ,   &  x \in [0, a],  \\
& \\
  D_a(a; \mu) + (x - a), &  x >a. 
\end{array}
  \right.
\end{equation}
In general, $D_a$ does not satisfy the smooth-fit condition at the boundary $a$. 
For any $x \geq 0$, the mapping $a \mapsto D_a(x; \mu)$ is maximized at the optimal policy $a   = \ba$ and  $D_{\ba}(x; \mu) = \Va(x)$. 
 
\subsection{Literature review}  
Both \cite{rd} and \cite{rs} allow a $(\mu, \sigma)$-pair among $\{(\mu_i, \sigma_i):  i = 1, 2, \dots, n\}$ to be part of the control, and \cite{rs} found which pair to use at any given value of $X(t)$.  The solution gave rise to some surprising results, for example that if the company is nearly bankrupt then it should be very conservative and use the $(\mu_i,\sigma_i)$-pair with the smallest $\sigma_i$ which seems to be paradoxical to many economists; see the work of \cite{sheth}.
For simplicity, in discussing the present question we will limit the company to only one corporate direction, i.e., $n = 1$. 

Many variants of the Radner--Shepp model have been proposed in the literature~\citep{taksar2000optimal, avanzi2009strategies, albrecher2009optimality}. \cite{decamps2007optimal} extended the Radner--Shepp model to include a singular control process representing an investment, and recently, \cite{de2017dividend} found the optimal policy for the finite horizon case. For optimal dividend distribution problems with general diffusion models, see \cite{paulsen2003optimal, paulsen2008optimal}.
It should be noted that for insurance companies, this dividend distribution problem also involves finding an optimal reinsurance policy, which gives rise to another control component that can affect both the profit rate (drift) and the riskiness (volatility) of the underlying fortune process; see, for example, \cite{taksar1998optimal, jgaard1999controlling, asmussen2000optimal, choulli2003diffusion, he2008optimal2}.  

Another important generalization of the dividend problem is to incorporate capital injections, which may be in the form of equity issuance. There is a vast literature on this topic; see, among  others, \cite{sethi2002optimal, lokka2008optimal, he2008optimal, kulenko2008optimal, yao2011optimal, zhou2012optimal, zhu2016optimal, chen2016optimal, schmidli2017capital, lindensjo2019optimal}. For a recent advance in the general theory, see~\cite{alvarez2018class}.  The model we will propose in the present paper allows fiscal stimulus in the form of a loan from the government. 
However, in all the above references, the capital injection process is allowed to be singular and, in particular, be of barrier type; in the present paper, we restrict the maximum rate of government loans (and thus require that the capital injection process be continuous.)

\section{Calculation of the value function $\Vb(x; \mu, \mu^*)$}\label{sec4}
We now state the main result of this paper. For our value function defined in equation~\eqref{eq:def.Vb},  under the optimal control, the dividend payout process is still of barrier type, though the threshold is different from that of Radner--Shepp model, and the company continuously borrows at the maximum possible rate, $c= \mu^*-\mu$.

\begin{theorem}[verification]\label{th:verify}
Consider the free-boundary problem 
\begin{align*}
\left\{
\begin{array}{cc}
\cL v(x) = c,  \quad \quad & x \in [0, a], \\
v(0) = 0,   \quad \quad  & \\
v'(x) = 1,   \quad \quad  & x \in [a, \infty), \\
v''(x) = 0,  \quad \quad & x \in [a, \infty), 
\end{array}
\right.
\end{align*}
where both $a$ and $v$ are unknown, $c = \mu^*-\mu \geq 0$ and 
\begin{equation*}\label{eq:def.L}
\cL = -r  + \mu^* \dfrac{\partial}{\partial x} + \dfrac{\sigma^2}{2} \dfrac{\partial^2}{\partial x^2} . 
\end{equation*}
Let $\hat{V} \in C^2$ and $\hat{a} \geq 0$ be the solution to this  problem. Then $\hv(x) = \Vb(x; \mu, \mu^*)$, the value function defined in equation ~\eqref{eq:def.Vb}.  
The associated optimal policy $(\hZ_+, \hZ_-)$ is   
\begin{equation*}
\hat{Z}_+(t) = \BP_{\ha}(t; \mu^*) = \sup_{0 \leq s \leq t}   ( x + \mu^* s + \sigma W(s) - \hat{a} )^+,  \quad \hat{Z}_-(t) = ct. 
\end{equation*}
\end{theorem}

\begin{proof}
The proof is given in Section~\ref{sec:proof}. 
\end{proof}

Next, we show that the solution to the  free-boundary problem given in Theorem~\ref{th:verify} always exists. 
We first notice that $\cL v = c$ is just a second-order linear equation with constant coefficients. Hence, standard differential equation results yield that for $x \in [0, \ha]$,  $\hv(x)$ is of the form  
\begin{align*}
-\frac{c}{r} + A_+  e^{\gamma^*_+ x} + A_- e^{\gamma^*_- x},
\end{align*}
where  constants $\gamma^*_+, \gamma^*_-$ are obtained by solving $\sigma^2  \gamma^2 / 2 + \mu^* \gamma - r = 0$, 
\begin{equation*} 
\gamma^*_\pm = \frac{-\mu^*\pm \sqrt{(\mu^*)^2 + 2r\sigma^2}}{\sigma^2}. 
\end{equation*}
Using  boundary conditions $\hv'(\ha) = 1$ and $\hv''(\ha) = 0$, we find that 
\begin{align*}
A_+ = \frac{-\gamma^*_- e^{-\gamma^*_+{\hat{a}}}}{\gamma^*_+(\gamma^*_+ - \gamma^*_-)}, \quad 
A_- = \frac{\gamma^*_+ e^{-\gamma^*_-{\hat{a}}}}{\gamma^*_- (\gamma^*_+ - \gamma^*_-)}.  
\end{align*} 
Since $\hv$ is linear on $ [\ha, \infty)$,  the solution can be written as  
\begin{align}
\hv(x) = \; & A_+ e^{\gamma^*_+ x} + A_- e^{\gamma^*_- x} -  (c / r),  \quad \;  x \in[0, \ha], \label{eq:sol.v1} \\
\hv(x) =\; &  \hv(\ha)+  (x - \ha), \quad \quad \quad\quad \quad \quad     x \in [\ha, \infty).  \label{eq:sol.v2}
\end{align}
Finally, the boundary condition $\hv(0) = 0$ implies that $\ha$ must satisfy 
\begin{equation} \label{impfunc}
f(\ha) = 0, \; \text{ where } f(a; \mu, \mu^*) = \frac{(-\gamma^*_-) e^{-\gamma^*_+ a}}{(\gamma^*_+)(\gamma^*_+ - \gamma^*_-)} + \frac{(\gamma^*_+) e^{-\gamma^*_- a}}{ \gamma^*_- (\gamma^*_+ - \gamma^*_-)} - \frac{c}{r}. 
\end{equation}

We denote the solution to the equation $f(a; \mu, \mu^*) = 0$  by $\ha = \aopt ( \mu, \mu^*)$ to emphasize its dependence on $\mu$ and $\mu^*$.
In Proposition~\ref{prop:exist}, we prove that such a solution always exists and is unique given that $\mu > 0$ and $\mu^* \geq \mu$. Hence, by solving $f(a) = 0$ we obtain  the optimal reflection barrier $\ha$, but, unlike in the Radner--Shepp model, it does not have a closed-form expression. 
The only exception is the special case $\mu^* = \mu$,  where we have $\aopt (\mu, \mu) = \ba$, the latter of which is as defined in equation~\eqref{eq:def.a1}. 

\begin{proposition}\label{prop:exist}
Assume $\mu > 0$ and $\mu^* \geq \mu$.   The free-boundary problem in Theorem~\ref{th:verify} has a unique solution $(\hv, \ha)$ such that $\hv \in C^2$ and $\ha > 0$. 
\end{proposition}

\begin{proof}
It suffices to show that $f(a; \mu, \mu^*) = 0$ has only one solution and it is positive, and then the rest follows from~\eqref{eq:sol.v1} and~\eqref{eq:sol.v2}.  Using~\eqref{impfunc}, it is straightforward to verify that $f$ is monotone decreasing,  $f(0) = \mu/r > 0$, and $f(\infty) = -\infty$.  So there is one and only one root on $(0, \infty)$. 
\end{proof}

From~\eqref{eq:vs.ineq}, we already know that   by borrowing money from the government, the company has an expected net value at least as large as in the Radner--Shepp model without government loans. 
Now we prove another key result of this work: as long as the optimal policy is used, the net value of the company is strictly larger; furthermore, the more money a company can borrow from the government, the larger net value it has. 

\begin{theorem}\label{th:strict}
Consider $x > 0$ and two pairs of  parameters $(\mu_1, \mu^*_1)$ and $(\mu_2, \mu^*_2)$. 
\begin{enumerate}[(i)]
\item If $\mu_1 = \mu_2 = \mu$ and $\mu^*_1 > \mu^*_2 \geq \mu$,  then  
$\Vb(x; \mu, \mu^*_1) > \Vb(x; \mu, \mu^*_2). $
\item If $\mu_1^* = \mu_2^* = \mu^*$ and $\mu_2 < \mu_1 \leq \mu^*$,  then 
$\Vb(x; \mu_1, \mu^* ) > \Vb(x; \mu_2, \mu^* ). $
\end{enumerate}
\end{theorem}
\begin{proof}
The proof for part (i) relies on the verification techniques used for proving Theorem~\ref{th:verify} and requires the result of Proposition~\ref{prop:ba}. 
Hence, the proof for this part is relegated to Section~\ref{sec:proof2}. 

To prove part (ii),  let $\ha_2 = \aopt ( \mu_2, \mu^* )$ be the solution to $f(a; \mu_2, \mu^*) = 0$.  Using the definition of the barrier payout process given in~\eqref{eq:def.Za}, the optimal control for a company characterized by the five-tuple $(x, \mu_2, \mu^*, \sigma, r)$ can be denoted by $( \BP_{\ha_2}(t; \mu^*), \;   (\mu^* - \mu_2) t)$. 
For another company characterized by   $(x, \mu_1, \mu^*, \sigma, r)$, the policy $( \BP_{\ha_2}(t; \mu^*), \;   (\mu^* - \mu_1) t)$   is admissible, which means to make dividend payments at reflection barrier $\ha_2$ but borrow money at rate $\mu^* - \mu_1 < \mu^* - \mu_2$. Hence,  
\begin{align*}
\Vb(x; \mu_1, \mu^*)  \geq \;& \sE_x  \left[ \int_{0-}^{\tau_0} e^{-rt}  d  \BP_{\ha_2}(t; \mu^*) -  \int_{0}^{\tau_0} (\mu^* - \mu_1) e^{-rt}  dt
 \right] \\
 = \;& \Vb(x; \mu_2, \mu^* ) + \sE_x \left[ \int_{0}^{\tau_0} (\mu_1 - \mu_2) e^{-rt}  dt \right]. 
\end{align*}
Given any $x > 0$, we have $\tau_0 > 0$ and thus part (ii) of the theorem follows from the assumption $\mu_1 > \mu_2$. 
\end{proof}

\section{Choice of the dividend payout barrier}\label{sec:barrier}

We  compare   $\ha =  \aopt(\mu, \mu^*)$ with two suboptimal choices:
\begin{equation*}\label{eq:def.2a}
 \ba   =  \aopt(\mu, \mu ), \quad \quad 
\sa  =  \aopt(\mu^*, \mu^* ). 
\end{equation*}
The threshold $\ba$ is the same as that defined in equation~\eqref{eq:def.a1}, which is the optimal threshold for the Radner--Shepp model where borrowing is not allowed. 
The threshold $\sa$ represents the greedy strategy of a firm that has original profit rate $\mu$ and borrows at rate $c = \mu^* - \mu$; the firm does not care about repaying the loan and thus chooses to use $\sa$ to maximize  the expected total dividend payouts (of course, $\sa$ is also the optimal threshold for the Radner--Shepp model  where the profit rate of the firm is $\mu^*$.)

We first prove  that $\ha $  is always less than the Radner--Shepp threshold $\ba$.  An immediate consequence is that since $\ha \in (0, \ba)$, $\ha$ can be computed numerically using a standard one-dimensional optimization algorithm.

\begin{proposition}\label{prop:ba}
Assume $\mu^* \geq \mu > 0$.  
Then $\ha = \aopt(\mu, \mu^* ) \le \aopt (\mu, \mu) = \ba$ where $\ba$ is the optimal threshold of the Radner--Shepp model. 
Further, for any $\mu^*_1 > \mu^*_2 \geq \mu$,   we have the strict inequality 
$ \aopt(\mu, \mu^*_1 )  < \aopt(\mu, \mu^*_2).$ 
\end{proposition}
\begin{proof}
Recall that in~\eqref{eq:va} we showed that $V( \aopt(\mu, \mu); \mu, \mu)    = \mu / r$. This is actually a special case of the identity 
\begin{align*}
\Vb( \aopt(\mu, \mu^*); \mu, \mu^*)   = \frac{\mu}{r},  \quad \quad \text{ for } \mu^* \geq \mu, 
\end{align*}
which can be straightforwardly verified using the boundary conditions in Theorem~\ref{th:verify}.
By~\eqref{eq:vs.ineq},  $\Vb(x; \mu, \mu^*) \ge \Vb(x; \mu, \mu)$ for every $x$. 
Further, both $\Vb(x; \mu, \mu)$ and $\Vb(x; \mu, \mu^*)$ are monotone increasing in $x$. Hence,  $\Vb(\ba; \mu, \mu) = \Vb(\ha; \mu, \mu^*) = \mu/r$  implies that $\ha \leq \ba$. 

To prove the strict inequality, consider the function $f$ defined in~\eqref{impfunc} and the mapping $\mu^* \mapsto f'(a;  \mu, \mu^*)$ where $f'$ denotes the derivative with respect to $a$.  Routine but heavy calculation gives 
\begin{align*}
\dfrac{\partial f'(a; \mu, \mu^*)}{ \partial \mu^*} =  -\dfrac{r \sigma^2  e^{-\gamma_+^* a}}{ [ (\mu^*)^2 + 2 r \sigma^2 ]^{3/2}} 
\left\{ 1 + h(\mu^*) + e^{2 h (\mu^*) } (  h (\mu^*) - 1 )  \right\},
\end{align*}
where $h(\mu^*) = a \sigma^{-2} \sqrt{  (\mu^*)^2 + 2 r \sigma^2  }  $. By computing the first two derivatives, one can verify that $h \mapsto 1 + h + e^{2h}(h - 1) $ is always positive on $(0, \infty)$.  Hence, for any $a \geq 0$, $\mu^* \mapsto f'(a; \mu, \mu^*)$ is monotone decreasing.  Combining this with the facts that $f(0; \mu, \mu^*) = \mu/r > 0$ and $ f'(a; \mu, \mu^*) < 0$ for any $a \geq 0$, we conclude that $\mu^* \mapsto \aopt(\mu, \mu^*)$ is also strictly monotone decreasing. 
\end{proof}

We prove in Proposition \ref{prop:greedy} below that $\ha$ is also smaller than the greedy threshold $\sa$ given $\mu^* > \mu$. 
Hence, in the absence of being held accountable, the firm taking the loan will pay out dividends later, which might be surprising.

\begin{proposition}\label{prop:greedy}
Assume $\mu^* > \mu > 0$. 
Then $  \ha = \aopt(\mu, \mu^* ) < \aopt (\mu^*, \mu^* ) =  \sa $ where $\sa$ is the optimal threshold of the Radner--Shepp model with profit rate $\mu^*$. 
More generally, for any $\mu^* \geq \mu_1 > \mu_2$, we have 
$ \aopt(\mu_1, \mu^* ) > \aopt(\mu_2, \mu^* ). $
\end{proposition}   
\begin{proof}
We only need to prove the general claim since $\sa = \ha(\mu^*, \mu^*)$ is a special case. 
Recall that, for any $\mu \leq \mu^*$, $\aopt(\mu, \mu^*)$ is the solution to $f(a; \mu, \mu^*) = 0$ where $f$ is as defined in~\eqref{impfunc},   and   $f(a; \mu, \mu^*)$  is monotone decreasing in $a$.  Observe that in the expression for $f$,  only the last term $c = \mu^* - \mu$ depends on $\mu$. 
Since $\mu$ only changes the vertical shift, but not the shape, of the function $f(a; \mu, \mu^*)$, we conclude that $\aopt(\mu_1, \mu^* ) > \aopt(\mu_2, \mu^*)$ as  $ \mu^* - \mu_1 <  \mu^* - \mu_2.$ 
\end{proof}   
    
To gain further insights into the problem, here we give an alternative proof for the inequality $\ha \leq \sa$. 
For our model, the net value of a company is the difference between the total dividend payouts and the total loan cost (both time-discounted). 
We use $(\BP_a(t; \mu^*), \, ct)$  to denote  a policy that  always borrows money at the maximum rate $c$ and pays out whatever amount that exceeds a threshold $a > 0$. 
For such a policy, the expected net value can be written as 
\begin{equation}\label{eq:def.Va}
V_a(x; \mu, \mu^*) = D_a(x; \mu^*) - C_a(x; \mu, \mu^*)
\end{equation}
where $D_a$ (the total dividend payouts) is as defined in~\eqref{eq:def.D} and can be computed by~\eqref{eq:Da}, and the total loan cost is given by 
\begin{align*}
 C_a(x) = C_a (x; \mu,\mu^* ) =  \sE_x \int_0^{\tau_0(a, \mu^*)} c e^{-rt} dt. 
\end{align*}
Note that since the company borrows at the maximum rate, both   $D_a$ and $\tau_0$ (the bankruptcy time) do not depend on the original profit rate $\mu$. 
Since $\ha$ is the optimal barrier for the value function in equation~\eqref{eq:def.Vb}, we have 
\begin{align*}
 D_{\ha}(x; \mu^*) - C_{\ha}(x; \mu, \mu^*) = V_{\ha}(x) \geq V_{\sa} ( x) = D_{\sa}(x; \mu^*) - C_{\sa}(x; \mu, \mu^*). 
\end{align*}
On the other hand, $a \mapsto D_a(x; \mu^*)$  is maximized at $a = \sa$ since $\sa$ is optimal for the Radner--Shepp model with profit rate $\mu^*$. 
Thus,  $D_{\ha}(x; \mu^*) \leq D_{\sa} (x; \mu^*)$,  which further implies that $C_{\ha}(x) \leq C_{\sa}(x)$.  For fixed $\mu, \mu^*$, the mapping $a \mapsto C_a(x)$  is monotone increasing since a larger value of the barrier would imply a longer expected ``lifetime" of the company. Hence, we conclude that $\ha \leq \sa$. 

To compute the function $C_a$, note that  
\begin{equation}\label{eq:def.ga}
C_a(x; \mu, \mu^*) =  \dfrac{c}{r}\left\{ 1 - g_a(x; \mu^*) \right\}, \quad  \text{ where }  g_a(x;   \mu^*) = \sE_x \left[ e^{-r\tau_0(a, \mu^*)} \right]. 
\end{equation}
As shown in~\cite{shreve1984optimal}, for $x \in [0, a]$,  $g_a(x)$ is the solution to the differential equation $\cL g  = 0$ with boundary conditions $g'(a) = 0$ and $g(0) = 1$~\citep[c.f.][]{darling1953first, lehoczky1977formulas}. Straightforward calculation then yields that 
\begin{equation*}
 g_a(x;  \mu^* )   = \dfrac{   \gamma_-^*  e^{- \gamma_+^* (a - x)} - \gamma_+^*  e^{-\gamma_-^* (a - x)}    }{   \gamma_-^*  e^{- \gamma_+^* a} - \gamma_+^*  e^{-\gamma_-^* a} }, 
 \quad \quad x \in [0, a]. 
\end{equation*}
For $x > a$,  we have $g_a(x; \mu^*) = g_a(a; \mu^*)$ due to the initial dividend payment that forces $X(0) = a$. 
Hence, for each $a \geq 0$, we can explicitly compute the value of $V_a$ defined in~\eqref{eq:def.Va}, and the mapping $a \mapsto V^*_a(x)$ must be maximized at $\ha = \ha(\mu, \mu^*)$.  

\begin{figure}[htbp!]
\begin{center}
\includegraphics[width=0.8\linewidth]{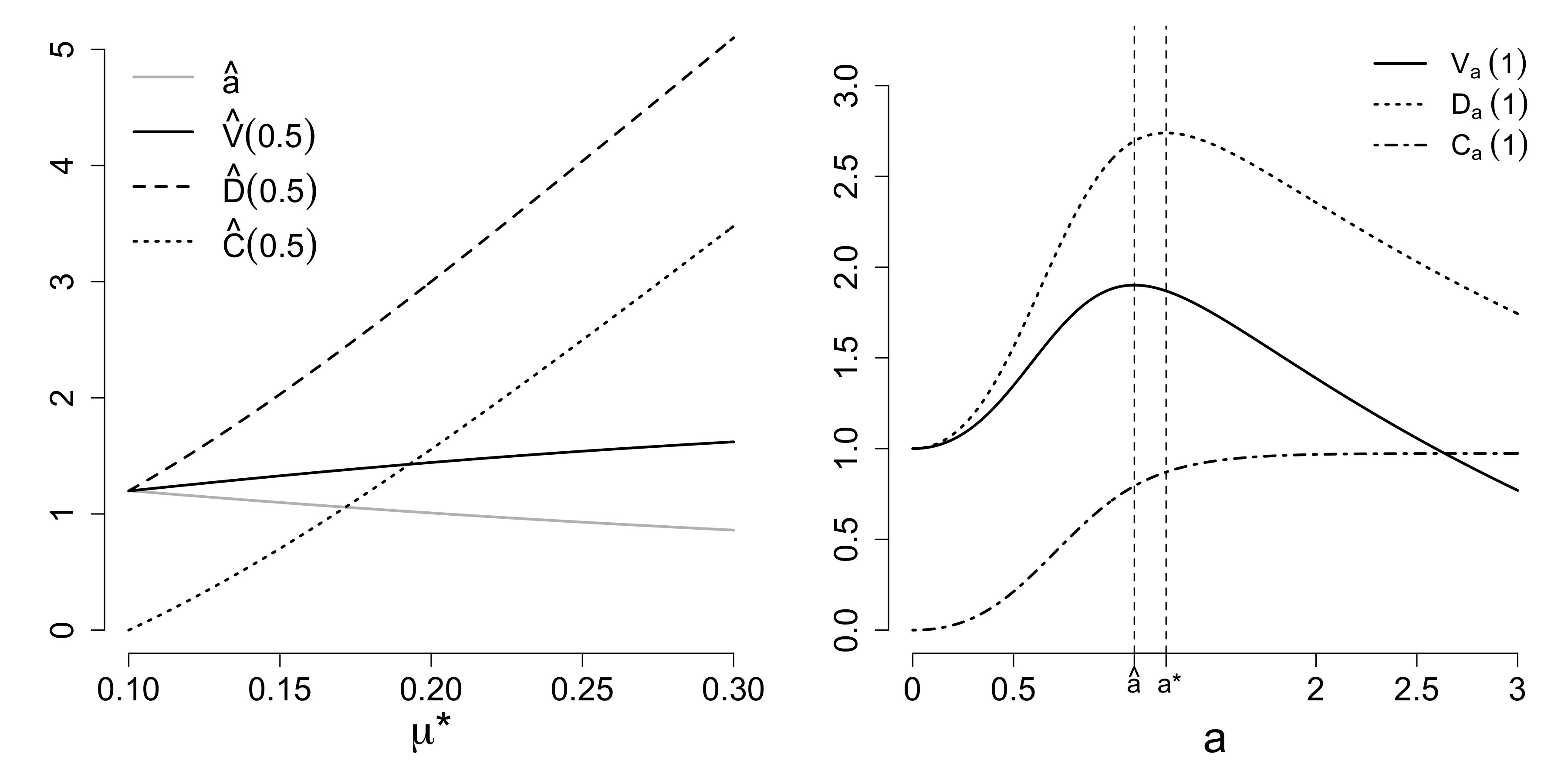}
\caption{Our model with $\mu = 0.1$, $r=0.05$ and $\sigma = 0.3$. 
In the left panel, we fix   $x = 0.5$, let $\mu^*$ range from $0.1$ to $0.3$, and plot 
$\ha = \aopt(\mu, \mu^*), \, \hat{V}(x) = V(x; \mu, \mu^*), \, \hat{D}(x) = D_{\ha}(x; \mu^*), \, \hat{C}(x) = C_{\ha}(x; \mu, \mu^*)$. 
Note that $\ba =\ha(\mu, \mu) = 1.200$. 
In the right, we fix $\mu^* = 0.15$, $x=1$, let barrier $a$ range from $0$ to $3$, and plot the expected net value $V_a(x) = V_a(x; \mu, \mu^*)$, the expected dividend payouts $D_a(x) = D_{a}(x; \mu^*)$, and the expected loan cost $C_a(x) = C_a(x; \mu, \mu^*)$.  Note that $V_a$ is maximized at $ \hat{a} = 1.099$ and  $D_{a}$ is maximized at $ \sa = 1.257.$} \label{fig2}
\end{center}
\end{figure}

\section{Discussion}\label{sec:disc}

\subsection{Analysis of the optimal payout policy}\label{sec:disc.opt}
We present a numerical example in Figure~\ref{fig2} illustrating the theoretical results proved in the previous section. 
By Theorem~\ref{th:strict},  when the  ``socially optimal" barrier $\hat{a} = \hat{a}(\mu, \mu^*)$ is used, the firm's expected net value increases monotonically as the size of the loan increases.  The left panel of Figure~\ref{fig2} shows the growth curve for a firm with original profit rate $\mu = 0.1$,  as $\mu^{*}$ increases from 0.1 (no government loan) to 0.3. 
The dividend payouts increase at a slightly faster rate than the cost of the loan, and thus the firm is able to add value by taking on the government loan, when cost is taken into account.  

By Proposition~\ref{prop:ba}, as the firm takes on an increasing amount of government loans, holding everything else constant, the barrier $\hat{a}$ actually decreases; see the gray line in the left panel of Figure~\ref{fig2}. 
In particular, we have $\ba > \ha$, where $\ba = \ha(\mu, \mu)$ is the optimal barrier when no money can be borrowed.  
From an economic perspective, this says that the firm will choose to pay dividends sooner if it receives more government funds.  
In the right panel of Figure 1,  $\mu^*$ is fixed at $0.15$ and we examine the three functions given in~\eqref{eq:def.Va}, $V_a,  D_a$ and $C_a$, for different barrier levels. 
The expected net value $V_a$  is maximized when the firm selects $\ha$, which incorporates repayment of the loan.  The function $D_a$ is maximized at $\sa$, the optimal barrier of a firm with profit rate $\mu^*$ but without access to government loans. As shown in Proposition~\ref{prop:greedy}, we always have $\ha < \sa$.

The observation that $\ha$ is less than either $\ba$ or $\sa$  reveals what economists have known for years about perverse financial incentives but resulting from a purely mathematical perspective within the constructs of our optimal control problem.     
When a firm is given money without oversight, or knows it will be bailed out, it may act more selfishly and recklessly (economists refer to this ``moral hazard".) This also justifies why, in reality, the government often requires, as a condition attached to stimulus packages or bailouts, that their loans must be paid back before companies can pay dividends to shareholders. The fiscal stimulus from the government in the form of a loan does boost the economy,  but with the caveat that there are some greedy incentives at work and therefore requires oversight.  

Nevertheless, we point out that $\ha < \ba$ does not mean that the firms may go bankruptcy faster when borrowing money. 
Actually, according to our numerical results (not shown here), we conjecture that for any given $\mu > r$  and $0 < x \leq \ha(\mu, \mu)$, the mapping $\mu^* \mapsto g_{\ha(\mu, \mu^*)}(x; \mu^*) \equiv \tilde{g}(\mu^*)$  is monotone decreasing on $[\mu, \infty)$, where $g$ is as given in~\eqref{eq:def.ga}. 
Since $\tilde{g}(\mu^*) = \sE_x \left[ e^{-r \tau_0} \right]$ where $\tau_0$ is the bankruptcy time of the firm using the barrier $\ha$,   this conjecture implies that the more money a firm can borrow, the longer lifetime it tends to have. 
Hence, government loans do improve the financial stability of firms in the sense that bankruptcy can be delayed. 

\subsection{Analysis of the greedy payout policy}\label{sec:disc.greedy}

If a company can borrow money at rate $\mu^* - \mu$ but does not care about repaying the loan, it would use the greedy  threshold $\sa$ since it maximizes the expected total dividend payouts. Numerical experiments were performed  to investigate the consequences of such greedy policies, from which we have made two interesting observations. 
 
\begin{figure}[htbp!]
\begin{center}
\includegraphics[width=0.5\linewidth]{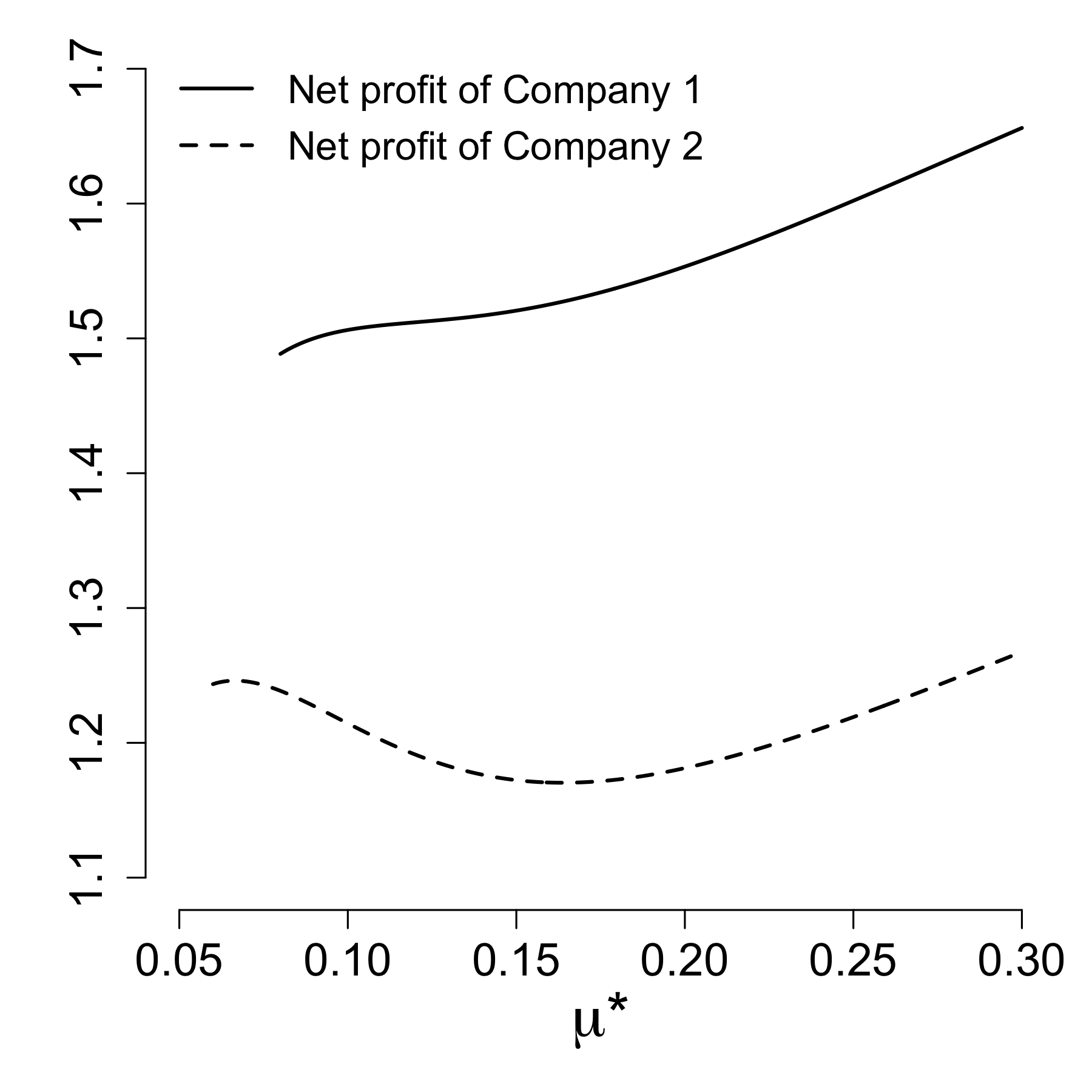}
\caption{Behavior of two ``greedy" companies with $r = 0.05$ and $x = 1$. We plot  $V_{\sa}(x; \mu, \mu^*)$ (the net value of a greedy policy) against the parameter $\mu^*$. The first company has original profit rate $\mu = 0.08$ 
and  the second has $\mu = 0.06$. Both have volatility $\sigma = 0.3$. 
} \label{fig3}
\end{center}
\end{figure}

First, the expected net value of a greedy company that does take the loan, $V_{\sa}(x; \mu, \mu^* )$, could be smaller than  the value of a company that does not take the loan, $V(x; \mu, \mu)$. 
That is, for a greedy company, the additional profit gained by borrowing money may not be able to even cover the loan cost! 
In Figure~\ref{fig3}, we give a numerical example of two companies. We fix $x = 1$, $r = 0.05$,  with an original profit rate $\mu = 0.08$ for Company 1 and $\mu = 0.06$ for Company 2. For Company 1 (solid line), $V_{\sa}(x; \mu, \mu^*)$ keeps increasing as $\mu^*$ increases,  which means that the more money it borrows, an even larger increase in value can be obtained  by using the greedy payout barrier $\sa$. However, for Company 2 (dashed line), when government loan is not allowed, it has $V(1; \mu, \mu) = 1.24$. But if $\mu^* = 0.16$, which means to borrow at rate $0.1$, $V_{\sa}(1; \mu, \mu^*)$ drops to $1.17$.  
Therefore, for Company 2,  its greedy payout policy is {\em socially undesirable}. 

We also find that $D_{\sa}(x; \mu^*)$ is not necessarily greater than or equal to $C_{\sa}(x; \mu, \mu^*)$; that is,  it could be dangerous of the government to lend money to some firms since they may not even be able to repay the loan (in expectation), though this happens rarely according to our numerics (only when both $\mu$ and $x$ are very small).  For example, let $\mu = 0.005,  \mu^* = 0.055,  \sigma = 0.1, r = 0.05$ and $x = 0.05$. One can compute that $\sa = 0.416$ and $\ha = 0.098$. Numerics show that as long as the payout barrier $a > 0.34$,  the net value, $V_a(x; \mu, \mu^*)$, would be negative.

The above observations  also lend support to our claim that government intervention may be needed in restricting the activities of the firms that receive stimulus until the loans are paid back.
One way to incorporate the government intervention in our mathematical model is to require that the  payout barrier has to be  within some pre-specified range, say $[a_{\rm{min}},  a_{\rm{max}}]$. 
The parameters $a_{\rm{min}}$ and  $a_{\rm{max}}$ should be calculated for each individual company separately in order to guarantee that the company is able to repay the loan.  
We note that a similar problem was considered by~\citet{paulsen2003optimal} assuming no capital injections or fiscal stimulus. 

\subsection{Extensions of our model}
We may interpret $r$  in our model as an exogenous parameter that  is charged by the government directly, 
as the government loans may not be  funneled through the banking system and can come in the form of a direct subsidy. 
One may also  consider a more general formulation of our problem with the value function~\citep[c.f.][]{lokka2008optimal}
\begin{equation*}
  \Vb(x; \mu,\mu^*, \beta) = \sup_{ (Z_+, Z_-) \in \cA } \sE_x    \int_{0-}^{\tau_0} e^{-rt} [ dZ_+(t) -  \beta dZ_-(t) ],  
\end{equation*}  
where $\beta \geq 1$ represents the proportional cost of borrowing money from the government.  When $\beta = 1$, there is no additional cost and the value function reduces to the one defined in equation ~\eqref{eq:def.Vb}. 
However, one can check that our argument cannot be straightforwardly extended to the case of $\beta > 1$, though it can be used for the case of $\beta < 1$, the case where we have some proportional cost of taking the dividends (see Section~\ref{sec:extension}.)
We leave the case of $\beta > 1$ to future study.
 
\section{Proofs}\label{sec:proofs}

\subsection{Proof for Theorem~\ref{th:verify}}\label{sec:proof}
The verification consists of two steps. First, we need to check that $\hv(x)$ is indeed the expected net value of the firm when we apply the candidate optimal control function $(\hat{Z}_+, \hat{Z}_-)$.  Second, we need to prove that no other policy can do better.  
The latter requires the following lemma. 

\begin{lemma}\label{lm:hv}
The solution $\hv$ given in~\eqref{eq:sol.v1} and~\eqref{eq:sol.v2} satisfies  $ \hv'(x) \geq  1 $ and  $\cL \hv(x) \leq  c$ for any $x \geq 0$. 
Consequently,  the following Hamilton-Jacobi-Bellman equation holds: 
\begin{equation*}\label{eq:hjb}
   \max\{  \cL \hv(x) - c, \, 1 -  \hv'(x) \} = 0,  \quad \quad \forall \, x \geq 0. 
\end{equation*}
\end{lemma}
\begin{proof}

We first show that $\hv''(x) \leq 0$ for any $x \geq 0$.  By the expression of $\hv$ given in~\eqref{eq:sol.v1} and~\eqref{eq:sol.v2}, this is equivalent to proving that, for any $x \in [0, \ha]$, 
\begin{equation}\label{eq:lm1}
A_+( \gamma_+^*)^2 e^{ (\gamma_+^* - \gamma_-^*) x} \leq - A_- (\gamma_-^*)^2. 
\end{equation}
Since by definition $\gamma_+^* > 0$ and $ \gamma_-^* < 0$, we only need show~\eqref{eq:lm1} holds true for $x = \ha$. But we already know that $V''(\ha) = 0$ and thus $A_+( \gamma_+^*)^2 e^{ (\gamma_+^* - \gamma_-^*) \ha} = - A_- (\gamma_-^*)^2.$ Hence we conclude $\hv''(x) \leq 0$.

Since $\hv'(x) = 1$ for any $x \geq \ha$, the non-positivity of $\hv''$ implies that  $\hv'(x) \geq 1$ for any $x \geq 0$. 
To prove $\cL \hv(x) \leq  c$, notice that $\hv(x) > \hv(\ha), \, \hv'(x) = \hv'(\ha),  \, \hv''(x) = \hv''(\ha)$ for any $x > \ha$. The claim then follows from the condition that $\cL \hv(\ha) = c$.
\end{proof}

\begin{remark}\label{rmk:smooth}
The smooth-fit condition, $\hv''(\ha) = 0$, is critical in the above proof.  
Assume all the other conditions in Theorem~\ref{th:verify} are satisfied. 
Since $\hv'(x) = 1$ for all $x \geq \ha$, $\hv''(\ha +) = 0$. 
If $\hv''(\ha-) > 0$, then $\hv'(\ha - \epsilon) < 1$  for some $\epsilon > 0$ since $\hv'(\ha) = 1$. 
If $\hv''(\ha -) < 0$, one can show that $\cL \hv(\ha + ) > c,$ since there is a jump increase in $\hv''(x)$ at $x = \ha$ and $\cL \hv(\ha -) = c$. 
Therefore, the smooth-fit condition is necessary for the Hamilton-Jacobi-Bellman equation to hold true. 
\end{remark}

Now we present our verification proof. 

\medskip 

\textbf{Step 1. }
We use $\hX$ to denote the cash reserve process when the candidate optimal control $(\hZ_+, \hZ_-)$ is applied, and let $\tau_0 = \tau_0^{\hX}$ be the time of bankruptcy.
First, consider the case $x \in [0, \ha]$.   The process $\hZ_+(t) = \BP_{\ha}(t; \mu^*)$  is just (a multiple of) the local time of the process $\hX$ at   level $\ha$~\citep[Chap. 3.6]{karatzas1998brownian}. 
Define $\alpha_t =  \brt  \wedge t$ and consider the process $e^{- r \alpha_t} \hv ( \hX ( \alpha_t )  ) $. By It\^o's formula,  
\begin{align*}
& e^{- r \alpha_t} \hv ( \hX ( \alpha_t) )   \\
=\;& \hv( x ) + \int_0^{\alpha_t}   e^{-rs}  \cL  \hv(\hX(s))ds  + \int_0^{\alpha_t}   e^{-r s } \hv'(\hX (s) )  \left[  \sigma dW(s) - d \hZ_+(s)  \right]. 
\end{align*}
For any $s \in [0, \brt]$, we always have $\hX(s) \in [0, \ha]$, and thus  $\cL \hv(\hX(s)) = c$ and $\hv'(\hX(s))$ stays bounded.  Hence, the integral with respect to $dW$ is zero.  Taking expectations on both sides we obtain
\begin{equation}\label{eq:exp.stop}
\sE_x \, e^{- r \alpha_t } \hv ( \hX (\alpha_t)) 
= \hv( x ) +   \sE_x  \int_0^{\alpha_t}   e^{-rs}  \left[ c \, ds  -     d \hZ_+(s)  \right],  
\end{equation}
where we have also used the fact that $\hv' (\ha) = 1$, and $ d\hZ_+ (t) = 0$ if $\hX(t) \neq \ha$. 
We now claim that 
\begin{equation}\label{eq:lim.zero}
\lim\limits_{t \rightarrow \infty } \sE_x \, e^{- r \alpha_t } \hv ( \hX (\alpha_t) ) = 0. 
\end{equation}
If bankruptcy happens, i.e., $\brt < \infty$, then $\hv(\hX(t)) = 0$ for any $t \geq \brt$;  if $\brt = \infty$, $\hv(\hX(t))$ always stays bounded and $e^{-r \alpha_t } \rightarrow 0$, from which~\eqref{eq:lim.zero} follows.  
For the integral on the right hand-side of~\eqref{eq:exp.stop},  we have 
\begin{align*}
\lim_{t \rightarrow \infty } \sE_x  \int_0^{\alpha_t}  c  e^{-rs}  ds  = \; & \sE_x \int_{0-}^{\tau_0^{\hX}}  c e^{- rs} ds  \leq \frac{c}{r},   \\ 
\lim_{t \rightarrow \infty } \sE_x  \int_0^{\alpha_t}   e^{-rs} d \hZ_+(s) = \; & \sE_x \int_{0-}^{\tau_0^{\hX}}  c e^{- rs} ds,  
\end{align*}
by monotone convergence theorem. Hence, letting $t \rightarrow \infty$ in~\eqref{eq:exp.stop}, we obtain  
\begin{equation}\label{eq:def.hv}
\hv(x) =     \sE_x \int_{0-}^{\tau_0^{\hX}}  e^{-rt} \left[   d\hZ_+(t) -  d\hZ_- (t) \right]. 
\end{equation}
In particular, we have established the equality~\eqref{eq:def.hv} for $x = \ha$, which, together with the expression of  $\hv$ given in~\eqref{eq:sol.v2}, can be used to show that~\eqref{eq:def.hv} also holds  for $x \in (\ha, \infty)$. 
 
\begin{remark}\label{rmk:va}
The smooth-fit condition,  $\hv''(\ha) = 0$,  is not used in Step 1. 
One can  verify that the function $V_a$ given in~\eqref{eq:def.Va} is the solution to the ordinary differential equation 
$  \cL v(x) = c$ for $x \in [0, a],$   with boundary conditions $v(0) = 0,  v'(a) = 1.$  Hence, the above argument also proves that  $V_a$   is indeed the expected net value of the policy $( \BP_a(t; \mu^*), \, (\mu^* - \mu)t)$. 
\end{remark}

\textbf{Step 2. }
For any admissible policy $Z = (Z_+, Z_-)$,  let  
$$V_Z(x ) =    \sE_x \int_{0-}^{\tau_0^X }  e^{-rt} \left[   dZ_+(t) -  dZ_- (t) \right]$$
be the expected net value of the firm.  Note that both $X$ and $\tau_0^X$ now depend on $(Z_+, Z_-)$, though this is not indicated explicitly in the notation. We need to show that $V_Z(x) \leq \hv(x).$
By assumption, $Z_-$ is a continuous process but $Z_+$ is not necessarily so. Therefore, we let $Z_+^c$ denote the continuous part of $Z_+$. 
The dynamics of $X$ is given by~\eqref{eq:def.XX}. Applying It\^o's formula to $\hat{V}(X_t)$, we obtain
\begin{equation}\label{eq:ineq.v}
 \hv( x )    
=   -  \int_0^{t \wedge \tau_0^X }  \sigma e^{-r s } \hv'(X(s) ) dW(s)
+ I_1 + I_2 - I_3 + I_4, 
\end{equation}
where 
\begin{equation}\label{eq:def.I}
\begin{aligned}
 I_1 = \;&  \int_0^{t \wedge \tau_0^X }   e^{-rs}  \left[  \hv'(X(s) ) ( -  d Z_-(s) + c \, ds  )  - \cL  \hv(X(s))ds  \right],  \\
I_2 = \; &  \int_0^{t \wedge \tau_0^X }   e^{-r s } \hv'(X(s) )  dZ_+^c(s)  , \\
I_3 = \;&  \sum\limits_{0 \leq  s  \leq (t \wedge \tau_0^X )}  e^{-r s} \left\{ \hv(X(s)) - \hv(X(s - )) \right\}, \\
I_4 = \;& e^{- r ( t \wedge \tau_0^X )} \hv ( X(t \wedge \tau_0^X) ).  
\end{aligned}
\end{equation}
From Lemma~\ref{lm:hv}, we have $\hv'(x) \geq 1$ and $\cL \hv(x) \leq c$. 
Moreover, since $c$ is the maximum rate of loans, we have $-  d Z_-(s) + c\, ds  \geq 0$. We then obtain that  
\begin{align*}
I_1  \geq   \int_0^{t \wedge \tau_0^X }     e^{-rs}  \left[  -   d Z_-(s) + c \, ds -  c \, ds \right]  =   - \int_0^{t \wedge \tau_0^X }     e^{-rs}     d Z_-(s)
\end{align*}
and 
\begin{align*}
I_2  \geq  \int_0^{t \wedge \tau_0^X }   e^{-r s }    dZ_+^c(s). 
\end{align*}
Further, $\hv'(x) \geq  1$ also implies that 
\begin{align*}
I_3   \leq  - \sum\limits_{0 \leq  s  \leq (t \wedge \tau_0^X )}  e^{-r s}  \Delta Z_+(s), 
\end{align*}
where we have used the fact that $X(s)  - X(s - ) = -( Z_+(s) - Z_+(s-) ).$
Taking expectations on both sides of~\eqref{eq:ineq.v} and using the boundedness of $\hv'$, we get 
\begin{align*}
\hv(x) 
\geq \; &  \sE_x \, \int_{0-}^{t \wedge \tau_0^X}   e^{-r s }  \left[ dZ_+(s)  -  d Z_-(s)  \right],
\end{align*}
since clearly $I_4 \geq 0$.  
Finally, by letting $t \rightarrow \infty$, applying monotone convergence theorem and noting that $\int_0^\infty e^{-rs} dZ_-(s) < \infty$, we obtain  
\begin{align*}
\hv(x)  \geq \; & \lim\limits_{t \rightarrow \infty}  \sE_x \, \int_{0-}^{t \wedge \tau_0^X}   e^{-r s }  \left[ dZ_+(s)  -  d Z_-(s)  \right] \\
= \;& \sE_x \, \int_{0-}^{ \tau_0^X}   e^{-r s }  \left[ dZ_+(s)  -  d Z_-(s)  \right] = V_Z(x), 
\end{align*}
 which completes the proof. 
 
\subsection{Proof for Theorem~\ref{th:strict} (i)} \label{sec:proof2} 
Let $b = \ha (\mu, \mu^*_2 )$. 
The value function $V(x; \mu, \mu^*_2 ) $   is attained by the policy, $\hZ_2 = ( \BP_b(t; \mu^*_2), \, (\mu^*_2 - \mu)t )$, where money is borrowed at rate $\mu^*_2 - \mu$ and  dividend payments are made at the barrier $b$. 
 Clearly, this policy is also admissible to a company characterized by $(x, \mu, \mu^*_1,  \sigma, r)$ since $\mu_1^* > \mu_2^*$, and we want to show that for any $x > 0$, 
\begin{align*}
\Vb(x; \mu, \mu^*_1 )  > \Vb (  x; \mu, \mu^*_2)  = V_{\hZ_2}(x; \mu, \mu^*_1). 
\end{align*}
Note that in the above verification proof, we have already shown $\Vb(x; \mu, \mu^*_1 ) \geq V_{\hZ_2}(x; \mu, \mu^*_1)$. 
Let $\hX_2(t) = x + \mu t + \sigma W(t)  +  d  \BP_b(t; \mu^*_2) -  (\mu^*_2 - \mu)t  $ denote the cash reserve process when the sub-optimal control $\hat{Z}_2$ is applied. 
Define 
\begin{align*}
\mathcal{L}_1^* = -r  + \mu^*_1 \dfrac{\partial}{\partial x} + \dfrac{\sigma^2}{2} \dfrac{\partial^2}{\partial x^2}. 
\end{align*}
By an argument wholly analogous to the Step 2 in Section~\ref{sec:proof}, when $\hat{Z}_2$ is applied, we have 
\begin{align*}
\Vb(x; \mu, \mu^*_1 )  \geq  \sE_x \, \left(  I_1 + I_2 - I_3  \right),
\end{align*}
where  
\begin{align*}
I_1  = \; & \int_0^{t \wedge \tau_0^{\hX_2} }   e^{-rs}  \left[  \Vb'( \hX_2(s); \mu, \mu^*_1 )   (\mu_1^*  - \mu_2^*) ds   - \cL_1  \Vb( \hX_2(s); \mu, \mu^*_1 )ds  \right],  \\ 
I_2 - I_3 \geq \;&  \int_{0-}^{t \wedge \tau_0^{\hX_2} }   e^{-r s }    d \BP_b (s; \mu^*_2). 
\end{align*}
To check that the above expression for $I_1$ agrees with~\eqref{eq:def.I}, note that for  $ Z =   \hat{Z}_2$ and $c = \mu_1^* - \mu $, we have 
\begin{align*}
   -  d Z_-(s) + c  ds  =  - (\mu^*_2 - \mu) ds + (\mu_1^* - \mu) ds   =  (\mu_1^*  - \mu_2^*) ds. 
\end{align*}

To show  that $\hZ_2$ is strictly sub-optimal, we need a slightly finer argument than that used to prove Theorem~\ref{th:verify}. 
Write $V_1 (x) = \Vb(x; \mu, \mu^*_1)$ to simplify the notation. 
Recall that $V_1(x ) $ is the solution to the free-boundary problem described in Theorem~\ref{th:verify} and  $\ha = \ha( \mu, \mu^*_1 )$ is strictly smaller than $b = \ha (\mu, \mu^*_2 )$ by Proposition~\ref{prop:ba}. Using Lemma~\ref{lm:hv}, it is then straightforward to verify that 
\begin{align*}
\begin{array}{cl}
 \mathcal{L}_1^* V_1 (x   ) < \mu_1^* - \mu ,  \quad  &  \forall \, x \in  ( \ha, b ] ,  \\
 V_1'(x  )  > 1,  \quad  &  \forall \, x \in  (0, \ha) , \\
  V_1'(x  )  = 1,  \quad  &  \forall \, x  \in [\ha, b].  
\end{array}
\end{align*} 
Further, the assumption $\mu_1^*  - \mu_2^* > 0$  implies that for any $x \in  (0, \ha) \cup ( \ha, b ]$, 
\begin{equation}\label{eq:ineq.v1}
\psi(x) =  V_1'( x   )  (\mu_1^*  - \mu_2^*)    - \cL_1  V_1( x  ) +  (\mu^*_2 - \mu)  > 0, 
\end{equation} 
and  $\psi(x) = 0$ if $x = \ha$. 
For any set $A \subseteq [0, \infty)$, define 
\begin{align*}
I_1(A) = \int_A   e^{-rs}  \left[  V_1'( \hX_2(s)  )   (\mu_1^*  - \mu_2^*) ds   - \cL_1  V_1( \hX_2(s)  )ds  \right]. 
\end{align*}
Choose an arbitrary $\delta > 0$, and let $A_0^\delta = [0,   \delta \wedge \tau_0^{\hX_2}]$ and $A_\delta^t  = ( \delta \wedge \tau_0^{\hX_2},  t \wedge \tau_0^{\hX_2}]$.  
For any $t \geq \delta$, we have $I_1 = I_1 (  [0,    t \wedge \tau_0^{\hX_2}] ) = I_1(A_0^\delta ) + I_1(A_\delta^t)$.  Consider 
\begin{align*}
 I_1(A_0^\delta )  +  \int_0^{ \delta \wedge \tau_0^{\hX_2} }     e^{-rs}    (\mu^*_2 - \mu) ds 
=  \int_0^{ \delta \wedge \tau_0^{\hX_2} } e^{-rs} \psi ( \hX_2(s)  ) ds. 
\end{align*}
Since $\hX_2(t)$ is a reflected Brownian motion,  given any $\hX_2(0) = x > 0$,  we have, almost surely, the bankruptcy time $\tau_0^{\hX_2} > 0$ and the Lebesgue measure of the set $\{0 \leq t \leq   \tau_0:  \, \hX_2(t) = \ha(\mu, \mu^*_1) \}$ is zero. 
Recall that for a non-negative measurable function, its Lebesgue integral is zero if and only if the function is zero almost everywhere. It then follows from~\eqref{eq:ineq.v1} that 
\begin{align*}
\sE_x   \int_0^{ \delta \wedge \tau_0^{\hX_2} } e^{-rs} \psi ( \hX_2(s)  ) ds = c_\delta(x) > 0, \quad \forall \, x > 0, 
\end{align*}
where $ c_\delta(x)$ only depends on $x$.  
Using~\eqref{eq:ineq.v1} again, we find that  for any $t \geq \delta$, 
\begin{align*}
\sE_x ( I_1)      \geq  c_\delta(x)  -  \sE_x     \int_0^{ t  \wedge \tau_0^{\hX_2} }    e^{-rs}    (\mu^*_2 - \mu) ds, 
\end{align*}
which further implies that 
\begin{align*}
V (x )  \geq\;&   \sE_x (  I_1 + I_2 - I_3 )  \geq    c_\delta(x) +   \sE_x    \int_{0-}^{t \wedge \tau_0}   e^{-r s }     \left[  d  \BP_b(s; \mu^*_2) -   (\mu^*_2 - \mu) ds  \right]. 
\end{align*}
Letting $t \rightarrow \infty$,  we conclude that 
\begin{align*}
\Vb(x; \mu, \mu^*_1 )  \geq  c_\delta(x) +   \Vb(x; \mu, \mu^*_2 ) >  \Vb(x; \mu, \mu^*_2 ), \quad \quad \forall \, x > 0. 
\end{align*}
  
\subsection{Extensions with proportional costs}\label{sec:extension}

For $\beta \in (0, 1]$, define the value function $V_\beta$ by 
 \begin{equation}\label{eq:def.Vbeta}
  \Vb_\beta (x; \mu,\mu^* ) = \sup_{ (Z_+, Z_-) \in \cA } \sE_x    \int_{0-}^{\tau_0} e^{-rt} [ dZ_+(t) -  \beta dZ_-(t) ]. 
\end{equation} 
Observe that 
\begin{align*}
\beta^{-1} \,  \Vb_\beta (x; \mu,\mu^* ) =  \sup_{ (Z_+, Z_-) \in \cA } \sE_x    \int_{0-}^{\tau_0} e^{-rt} [ \beta^{-1} dZ_+(t) -    dZ_-(t) ]. 
\end{align*}
Hence, the problem~\eqref{eq:def.Vbeta} with $\beta < 1$ can be interpreted as an extension of the main problem defined in~\eqref{eq:def.Vb} where there is a proportional cost of taking the dividends. 
For this problem, the optimal policy is  to borrow the money at maximum rate $c = \mu^* - \mu$ and make dividend payouts at some barrier $\hat{a}_\beta$.  
Further, $(V_\beta, \hat{a}_\beta)$ is the solution to the following free-boundary problem 
\begin{align*}
\left\{
\begin{array}{cc}
\cL v(x) = \beta c,  \quad \quad & x \in [0, a], \\
v(0) = 0,   \quad \quad  & \\
v'(x) = 1,   \quad \quad  & x \in [a, \infty), \\
v''(x) = 0,  \quad \quad & x \in [a, \infty). 
\end{array}
\right.
\end{align*}
By the argument following Theorem~\ref{th:verify}, the value function $V_\beta$ can still be written in the form of~\eqref{eq:sol.v1} and~\eqref{eq:sol.v2}. To check the existence of $\hat{a}_\beta$, one just need to verify the following function has a unique positive solution, 
\begin{equation*} 
  f_\beta(a) = \frac{(-\gamma^*_-) e^{-\gamma^*_+ a}}{(\gamma^*_+)(\gamma^*_+ - \gamma^*_-)} + \frac{(\gamma^*_+) e^{-\gamma^*_- a}}{ \gamma^*_- (\gamma^*_+ - \gamma^*_-)} - \frac{\beta c}{r}. 
\end{equation*}
By Proposition~\ref{prop:exist}, $f_\beta$ is monotone decreasing to $-\infty$ and observe that $f_\beta(0) = [ (1 - \beta)\mu^* + \beta \mu ] / r> 0$.  Hence, $\hat{a}_\beta$ exists uniquely. 
Similarly to Lemma~\ref{lm:hv}, one can show that  $ \max\{  \cL V_\beta(x) - \beta c, \, 1 -  V'_\beta(x) \} = 0$  for all   $x \geq 0$. 
 
The verification proof  is almost the same as  in the case of $\beta = 1$. 
The only step that does not follow immediately from the proof in Section~\ref{sec:proof}
is how to bound the term $I_1$ defined in~\eqref{eq:def.I}. 
Note that since $V'_\beta \geq 1 \geq \beta$ , $\cL V_\beta \leq \beta c$ and $-  d Z_-(s) / ds + c   \geq 0$, we have 
\begin{align*}
I_1 = \;&  \int_0^{t \wedge \tau_0^X }   e^{-rs}  \left[  V'_\beta(X(s) ) ( -  d Z_-(s) + c \, ds  )  - \cL  V_\beta(X(s))ds  \right],  \\
\geq \;&  \int_0^{t \wedge \tau_0^X }   e^{-rs}  \left[ \beta ( -  d Z_-(s) + c \, ds  )  - \beta c \, d s\right] 
=  -  \int_0^{t \wedge \tau_0^X }   e^{-rs}  \beta d Z_-(s). 
\end{align*}
The rest then follows. 
 
\section*{Acknowledgments}
\noindent The first-named author is grateful to ARO-YIP-71636-MA, NSF DMS-1811936, and ONR N00014-18-1-2192 for their support of this research. We are grateful for two anonymous referees, whose reports have greatly helped to improve the quality of this manuscript. This is a signicantly revised version of a paper originally written by Shepp and Imerman with the title ``Is mathematics able to give insight into current questions in finance, economics and politics?'' The original version can be found at \url{https://arxiv.org/abs/1410.6084v1}.

\newpage

\section*{References}
\bibliographystyle{model5-names}
\bibliography{ref}

\begin{thebibliography}{35}
\expandafter\ifx\csname natexlab\endcsname\relax\def\natexlab#1{#1}\fi
\providecommand{\url}[1]{\texttt{#1}}
\providecommand{\href}[2]{#2}
\providecommand{\path}[1]{#1}
\providecommand{\DOIprefix}{doi:}
\providecommand{\ArXivprefix}{arXiv:}
\providecommand{\URLprefix}{URL: }
\providecommand{\Pubmedprefix}{pmid:}
\providecommand{\doi}[1]{\href{http://dx.doi.org/#1}{\path{#1}}}
\providecommand{\Pubmed}[1]{\href{pmid:#1}{\path{#1}}}
\providecommand{\bibinfo}[2]{#2}
\ifx\xfnm\relax \def\xfnm[#1]{\unskip,\space#1}\fi
\bibitem[{Albrecher \& Thonhauser(2009)}]{albrecher2009optimality}
\bibinfo{author}{Albrecher, H.}, \& \bibinfo{author}{Thonhauser, S.}
  (\bibinfo{year}{2009}).
\newblock \bibinfo{title}{Optimality results for dividend problems in
  insurance}.
\newblock {\it \bibinfo{journal}{RACSAM-Revista de la Real Academia de Ciencias
  Exactas, Fisicas y Naturales. Serie A. Matematicas}\/},  {\it
  \bibinfo{volume}{103}\/}, \bibinfo{pages}{295--320}.
\bibitem[{Alvarez(2018)}]{alvarez2018class}
\bibinfo{author}{Alvarez, L.} (\bibinfo{year}{2018}).
\newblock \bibinfo{title}{A class of solvable stationary singular stochastic
  control problems, e-print}.
\newblock {\it \bibinfo{journal}{arXiv preprint ArXiv:1803.03464}\/}, .
\bibitem[{Asmussen et~al.(2000)Asmussen, H{\o}jgaard \&
  Taksar}]{asmussen2000optimal}
\bibinfo{author}{Asmussen, S.}, \bibinfo{author}{H{\o}jgaard, B.}, \&
  \bibinfo{author}{Taksar, M.} (\bibinfo{year}{2000}).
\newblock \bibinfo{title}{Optimal risk control and dividend distribution
  policies. example of excess-of loss reinsurance for an insurance
  corporation}.
\newblock {\it \bibinfo{journal}{Finance and Stochastics}\/},  {\it
  \bibinfo{volume}{4}\/}, \bibinfo{pages}{299--324}.
\bibitem[{Asmussen \& Taksar(1997)}]{asmussen1997controlled}
\bibinfo{author}{Asmussen, S.}, \& \bibinfo{author}{Taksar, M.}
  (\bibinfo{year}{1997}).
\newblock \bibinfo{title}{Controlled diffusion models for optimal dividend
  payout}.
\newblock {\it \bibinfo{journal}{Insurance: Mathematics and Economics}\/},
  {\it \bibinfo{volume}{20}\/}, \bibinfo{pages}{1--15}.
\bibitem[{Avanzi(2009)}]{avanzi2009strategies}
\bibinfo{author}{Avanzi, B.} (\bibinfo{year}{2009}).
\newblock \bibinfo{title}{Strategies for dividend distribution: a review}.
\newblock {\it \bibinfo{journal}{North American Actuarial Journal}\/},  {\it
  \bibinfo{volume}{13}\/}, \bibinfo{pages}{217--251}.
\bibitem[{Chen et~al.(2016)Chen, Wang, Deng \& Zeng}]{chen2016optimal}
\bibinfo{author}{Chen, S.}, \bibinfo{author}{Wang, X.}, \bibinfo{author}{Deng,
  Y.}, \& \bibinfo{author}{Zeng, Y.} (\bibinfo{year}{2016}).
\newblock \bibinfo{title}{Optimal dividend-financing strategies in a dual risk
  model with time-inconsistent preferences}.
\newblock {\it \bibinfo{journal}{Insurance: Mathematics and Economics}\/},
  {\it \bibinfo{volume}{67}\/}, \bibinfo{pages}{27--37}.
\bibitem[{Choulli et~al.(2003)Choulli, Taksar \& Zhou}]{choulli2003diffusion}
\bibinfo{author}{Choulli, T.}, \bibinfo{author}{Taksar, M.}, \&
  \bibinfo{author}{Zhou, X.~Y.} (\bibinfo{year}{2003}).
\newblock \bibinfo{title}{A diffusion model for optimal dividend distribution
  for a company with constraints on risk control}.
\newblock {\it \bibinfo{journal}{SIAM Journal of Control and Optimization}\/},
  {\it \bibinfo{volume}{41}\/}, \bibinfo{pages}{1946--1979}.
\bibitem[{Darling \& Siegert(1953)}]{darling1953first}
\bibinfo{author}{Darling, D.~A.}, \& \bibinfo{author}{Siegert, A.}
  (\bibinfo{year}{1953}).
\newblock \bibinfo{title}{The first passage problem for a continuous markov
  process}.
\newblock {\it \bibinfo{journal}{The Annals of Mathematical Statistics}\/},
  (pp. \bibinfo{pages}{624--639}).
\bibitem[{De~Angelis \& Ekstr{\"o}m(2017)}]{de2017dividend}
\bibinfo{author}{De~Angelis, T.}, \& \bibinfo{author}{Ekstr{\"o}m, E.}
  (\bibinfo{year}{2017}).
\newblock \bibinfo{title}{The dividend problem with a finite horizon}.
\newblock {\it \bibinfo{journal}{The Annals of Applied Probability}\/},  {\it
  \bibinfo{volume}{27}\/}, \bibinfo{pages}{3525--3546}.
\bibitem[{D{\'e}camps \& Villeneuve(2007)}]{decamps2007optimal}
\bibinfo{author}{D{\'e}camps, J.-P.}, \& \bibinfo{author}{Villeneuve, S.}
  (\bibinfo{year}{2007}).
\newblock \bibinfo{title}{Optimal dividend policy and growth option}.
\newblock {\it \bibinfo{journal}{Finance and Stochastics}\/},  {\it
  \bibinfo{volume}{11}\/}, \bibinfo{pages}{3--27}.
\bibitem[{Dutta \& Radner(1999)}]{rd}
\bibinfo{author}{Dutta, P.~K.}, \& \bibinfo{author}{Radner, R.}
  (\bibinfo{year}{1999}).
\newblock \bibinfo{title}{Profit maximization and the market selection
  hypothesis}.
\newblock {\it \bibinfo{journal}{The Review of Economic Studies}\/},  {\it
  \bibinfo{volume}{66}\/}, \bibinfo{pages}{769--798}.
\bibitem[{He et~al.(2008)He, Hou \& Liang}]{he2008optimal2}
\bibinfo{author}{He, L.}, \bibinfo{author}{Hou, P.}, \& \bibinfo{author}{Liang,
  Z.} (\bibinfo{year}{2008}).
\newblock \bibinfo{title}{Optimal control of the insurance company with
  proportional reinsurance policy under solvency constraints}.
\newblock {\it \bibinfo{journal}{Insurance: Mathematics and Economics}\/},
  {\it \bibinfo{volume}{43}\/}, \bibinfo{pages}{474--479}.
\bibitem[{He \& Liang(2008)}]{he2008optimal}
\bibinfo{author}{He, L.}, \& \bibinfo{author}{Liang, Z.}
  (\bibinfo{year}{2008}).
\newblock \bibinfo{title}{Optimal financing and dividend control of the
  insurance company with proportional reinsurance policy}.
\newblock {\it \bibinfo{journal}{Insurance: Mathematics and Economics}\/},
  {\it \bibinfo{volume}{42}\/}, \bibinfo{pages}{976--983}.
\bibitem[{H{\o}jgaard \& Taksar(1999)}]{jgaard1999controlling}
\bibinfo{author}{H{\o}jgaard, B.}, \& \bibinfo{author}{Taksar, M.}
  (\bibinfo{year}{1999}).
\newblock \bibinfo{title}{Controlling risk exposure and dividends payout
  schemes: insurance company example}.
\newblock {\it \bibinfo{journal}{Mathematical Finance}\/},  {\it
  \bibinfo{volume}{9}\/}, \bibinfo{pages}{153--182}.
\bibitem[{Jeanblanc-Picqu{\'e} \& Shiryaev(1995)}]{jeanblanc1995optimization}
\bibinfo{author}{Jeanblanc-Picqu{\'e}, M.}, \& \bibinfo{author}{Shiryaev,
  A.~N.} (\bibinfo{year}{1995}).
\newblock \bibinfo{title}{Optimization of the flow of dividends}.
\newblock {\it \bibinfo{journal}{Russian Mathematical Surveys}\/},  {\it
  \bibinfo{volume}{50}\/}, \bibinfo{pages}{257}.
\bibitem[{Karatzas \& Shreve(1998)}]{karatzas1998brownian}
\bibinfo{author}{Karatzas, I.}, \& \bibinfo{author}{Shreve, S.~E.}
  (\bibinfo{year}{1998}).
\newblock \bibinfo{title}{Brownian motion}.
\newblock In {\it \bibinfo{booktitle}{Brownian Motion and Stochastic
  Calculus}\/} (pp. \bibinfo{pages}{47--127}).
\newblock \bibinfo{publisher}{Springer}.
\bibitem[{Kulenko \& Schmidli(2008)}]{kulenko2008optimal}
\bibinfo{author}{Kulenko, N.}, \& \bibinfo{author}{Schmidli, H.}
  (\bibinfo{year}{2008}).
\newblock \bibinfo{title}{Optimal dividend strategies in a cram{\'e}r--lundberg
  model with capital injections}.
\newblock {\it \bibinfo{journal}{Insurance: Mathematics and Economics}\/},
  {\it \bibinfo{volume}{43}\/}, \bibinfo{pages}{270--278}.
\bibitem[{Lehoczky(1977)}]{lehoczky1977formulas}
\bibinfo{author}{Lehoczky, J.~P.} (\bibinfo{year}{1977}).
\newblock \bibinfo{title}{Formulas for stopped diffusion processes with
  stopping times based on the maximum}.
\newblock {\it \bibinfo{journal}{The Annals of Probability}\/},  {\it
  \bibinfo{volume}{5}\/}, \bibinfo{pages}{601--607}.
\bibitem[{Lindensj{\"o} \& Lindskog(2019)}]{lindensjo2019optimal}
\bibinfo{author}{Lindensj{\"o}, K.}, \& \bibinfo{author}{Lindskog, F.}
  (\bibinfo{year}{2019}).
\newblock \bibinfo{title}{Optimal dividends and capital injection under
  dividend restrictions}.
\newblock {\it \bibinfo{journal}{arXiv preprint arXiv:1902.06294}\/}, .
\bibitem[{L{\o}kka \& Zervos(2008)}]{lokka2008optimal}
\bibinfo{author}{L{\o}kka, A.}, \& \bibinfo{author}{Zervos, M.}
  (\bibinfo{year}{2008}).
\newblock \bibinfo{title}{Optimal dividend and issuance of equity policies in
  the presence of proportional costs}.
\newblock {\it \bibinfo{journal}{Insurance: Mathematics and Economics}\/},
  {\it \bibinfo{volume}{42}\/}, \bibinfo{pages}{954--961}.
\bibitem[{Lucas(2016)}]{lucas}
\bibinfo{author}{Lucas, D.} (\bibinfo{year}{2016}).
\newblock \bibinfo{title}{Credit policy as fiscal policy}.
\newblock {\it \bibinfo{journal}{Brookings Papers on Economic Activity}\/},
  {\it \bibinfo{volume}{2016}\/}, \bibinfo{pages}{1--57}.
\bibitem[{Paulsen(2003)}]{paulsen2003optimal}
\bibinfo{author}{Paulsen, J.} (\bibinfo{year}{2003}).
\newblock \bibinfo{title}{Optimal dividend payouts for diffusions with solvency
  constraints}.
\newblock {\it \bibinfo{journal}{Finance and Stochastics}\/},  {\it
  \bibinfo{volume}{7}\/}, \bibinfo{pages}{457--473}.
\bibitem[{Paulsen(2008)}]{paulsen2008optimal}
\bibinfo{author}{Paulsen, J.} (\bibinfo{year}{2008}).
\newblock \bibinfo{title}{Optimal dividend payments and reinvestments of
  diffusion processes with both fixed and proportional costs}.
\newblock {\it \bibinfo{journal}{SIAM Journal on Control and Optimization}\/},
  {\it \bibinfo{volume}{47}\/}, \bibinfo{pages}{2201--2226}.
\bibitem[{Radner \& Shepp(1996)}]{rs}
\bibinfo{author}{Radner, R.}, \& \bibinfo{author}{Shepp, L.}
  (\bibinfo{year}{1996}).
\newblock \bibinfo{title}{Risk vs. profit potential: A model for corporate
  strategy}.
\newblock {\it \bibinfo{journal}{Journal of Economic Dynamics and Control}\/},
  {\it \bibinfo{volume}{20}\/}, \bibinfo{pages}{1373--1393}.
\bibitem[{Schmidli(2017)}]{schmidli2017capital}
\bibinfo{author}{Schmidli, H.} (\bibinfo{year}{2017}).
\newblock \bibinfo{title}{On capital injections and dividends with tax in a
  diffusion approximation}.
\newblock {\it \bibinfo{journal}{Scandinavian Actuarial Journal}\/},  {\it
  \bibinfo{volume}{2017}\/}, \bibinfo{pages}{751--760}.
\bibitem[{Sethi \& Taksar(2002)}]{sethi2002optimal}
\bibinfo{author}{Sethi, S.~P.}, \& \bibinfo{author}{Taksar, M.~I.}
  (\bibinfo{year}{2002}).
\newblock \bibinfo{title}{Optimal financing of a corporation subject to random
  returns}.
\newblock {\it \bibinfo{journal}{Mathematical Finance}\/},  {\it
  \bibinfo{volume}{12}\/}, \bibinfo{pages}{155--172}.
\bibitem[{Sheth et~al.(2011)Sheth, Shepp \& Palmon}]{sheth}
\bibinfo{author}{Sheth, A.}, \bibinfo{author}{Shepp, L.}, \&
  \bibinfo{author}{Palmon, O.} (\bibinfo{year}{2011}).
\newblock \bibinfo{title}{Risk-taking, financial distress and innovation}.
\newblock {\it \bibinfo{journal}{Academy of Business Journal: Special Issue on
  the Global Debt Crisis}\/},  {\it \bibinfo{volume}{2}\/},
  \bibinfo{pages}{5--18}.
\bibitem[{Shreve et~al.(1984)Shreve, Lehoczky \& Gaver}]{shreve1984optimal}
\bibinfo{author}{Shreve, S.~E.}, \bibinfo{author}{Lehoczky, J.~P.}, \&
  \bibinfo{author}{Gaver, D.~P.} (\bibinfo{year}{1984}).
\newblock \bibinfo{title}{Optimal consumption for general diffusions with
  absorbing and reflecting barriers}.
\newblock {\it \bibinfo{journal}{SIAM Journal on Control and Optimization}\/},
  {\it \bibinfo{volume}{22}\/}, \bibinfo{pages}{55--75}.
\bibitem[{Taksar(2000)}]{taksar2000optimal}
\bibinfo{author}{Taksar, M.~I.} (\bibinfo{year}{2000}).
\newblock \bibinfo{title}{Optimal risk and dividend distribution control models
  for an insurance company}.
\newblock {\it \bibinfo{journal}{Mathematical methods of operations
  research}\/},  {\it \bibinfo{volume}{51}\/}, \bibinfo{pages}{1--42}.
\bibitem[{Taksar \& Zhou(1998)}]{taksar1998optimal}
\bibinfo{author}{Taksar, M.~I.}, \& \bibinfo{author}{Zhou, X.~Y.}
  (\bibinfo{year}{1998}).
\newblock \bibinfo{title}{Optimal risk and dividend control for a company with
  a debt liability}.
\newblock {\it \bibinfo{journal}{Insurance: Mathematics and Economics}\/},
  {\it \bibinfo{volume}{22}\/}, \bibinfo{pages}{105--122}.
\bibitem[{Tanaka(1979)}]{tanaka1979}
\bibinfo{author}{Tanaka, H.} (\bibinfo{year}{1979}).
\newblock \bibinfo{title}{Stochastic differential equations with reflecting
  boundary condition in convex regions}.
\newblock {\it \bibinfo{journal}{Hiroshima Math. J.}\/},  {\it
  \bibinfo{volume}{9}\/}, \bibinfo{pages}{163--177}.
  \DOIprefix\doi{10.32917/hmj/1206135203}.
\bibitem[{Watanabe(1971)}]{watanabe1971stochastic}
\bibinfo{author}{Watanabe, S.} (\bibinfo{year}{1971}).
\newblock \bibinfo{title}{On stochastic differential equations for
  multi-dimensional diffusion processes with boundary conditions}.
\newblock {\it \bibinfo{journal}{Journal of Mathematics of Kyoto
  University}\/},  {\it \bibinfo{volume}{11}\/}, \bibinfo{pages}{169--180}.
\bibitem[{Yao et~al.(2011)Yao, Yang \& Wang}]{yao2011optimal}
\bibinfo{author}{Yao, D.}, \bibinfo{author}{Yang, H.}, \&
  \bibinfo{author}{Wang, R.} (\bibinfo{year}{2011}).
\newblock \bibinfo{title}{Optimal dividend and capital injection problem in the
  dual model with proportional and fixed transaction costs}.
\newblock {\it \bibinfo{journal}{European Journal of Operational Research}\/},
  {\it \bibinfo{volume}{211}\/}, \bibinfo{pages}{568--576}.
\bibitem[{Zhou \& Yuen(2012)}]{zhou2012optimal}
\bibinfo{author}{Zhou, M.}, \& \bibinfo{author}{Yuen, K.~C.}
  (\bibinfo{year}{2012}).
\newblock \bibinfo{title}{Optimal reinsurance and dividend for a diffusion
  model with capital injection: Variance premium principle}.
\newblock {\it \bibinfo{journal}{Economic Modelling}\/},  {\it
  \bibinfo{volume}{29}\/}, \bibinfo{pages}{198--207}.
\bibitem[{Zhu \& Yang(2016)}]{zhu2016optimal}
\bibinfo{author}{Zhu, J.}, \& \bibinfo{author}{Yang, H.}
  (\bibinfo{year}{2016}).
\newblock \bibinfo{title}{Optimal capital injection and dividend distribution
  for growth restricted diffusion models with bankruptcy}.
\newblock {\it \bibinfo{journal}{Insurance: Mathematics and Economics}\/},
  {\it \bibinfo{volume}{70}\/}, \bibinfo{pages}{259--271}.

\end{thebibliography}

\end{document}